\documentclass[aps,showpacs,twocolumn,pra,groupedaddress]{revtex4}
\usepackage{epsfig,amssymb,amsmath}
\usepackage[hypertex,linkcolor=red]{hyperref}
\def\comment#1{}
\def\labell#1{\label{#1}}
\def\togli#1{}

\begin{document} 
\title{The quantum mechanics of time travel through post-selected
  teleportation}
\author{Seth Lloyd$^{1}$, Lorenzo Maccone$^{1}$, Raul
  Garcia-Patron$^{1}$, Vittorio Giovannetti$^{2}$, Yutaka
  Shikano$^{1,3}$} \affiliation{$^{1}$xQIT,Massachusetts Institute of
  Technology, 77 Mass Ave, Cambridge MA.\\
  $^2$NEST-CNR-INFM \& Scuola Normale Superiore, Piazza dei Cavalieri
  7, I-56126, Pisa, Italy. \\ $^3$Dep. Physics, Tokyo Institute of
  Technology, 2-12-1 Oh-Okayama, Meguro, Tokyo, 152-8551, Japan.}

\begin{abstract}
  This paper discusses the quantum mechanics of closed timelike curves
  (CTCs) and of other potential methods for time travel.  We analyze a
  specific proposal for such quantum time travel, the quantum
  description of CTCs based on post-selected teleportation (P-CTCs).
  We compare the theory of P-CTCs to previously proposed quantum
  theories of time travel: the theory is physically inequivalent to
  Deutsch's theory of CTCs, but it is consistent with path-integral
  approaches (which are the best suited for analyzing quantum field
  theory in curved spacetime). We derive the dynamical equations that
  a chronology-respecting system interacting with a CTC will
  experience.  We discuss the possibility of time travel in the
  absence of general relativistic closed timelike curves, and
  investigate the implications of P-CTCs for enhancing the power of
  computation.
\end{abstract}
\pacs{03.67.-a,03.65.Ud,04.00.00,04.62.+v,04.60.-m} 
\maketitle

Einstein's theory of general relativity allows the existence of closed
timelike curves, paths through spacetime that, if followed, allow a
time traveler -- whether human being or elementary particle -- to
interact with her former self.  The possibility of such closed
timelike curves (CTCs) was pointed out by Kurt G\"odel~\cite{GODEL},
and a variety of spacetimes containing closed timelike curves have
been proposed ~\cite{Bonnor, Gott}.  Reconciling closed timelike
curves with quantum mechanics is a difficult problem that has been
addressed repeatedly, for example, using path integral
techniques~\cite{altri, politzer, boulware, hartle, STOCKUM,
  politzer2}.  This paper explores a particular version of closed
timelike curves based on combining quantum teleportation with
post-selection.  The resulting post-selected closed timelike curves
(P-CTCs) provide a self-consistent picture of the quantum mechanics of
time-travel.  P-CTCs offer a theory of closed timelike curves that is
physically inequivalent to other Hilbert-space based theories, e.g.,
that of Deutsch ~\cite{deutsch}.  As in all versions of time travel,
closed timelike curves embody apparent paradoxes, such as the
grandfather paradox, in which the time traveller inadvertently or on
purpose performs an action that causes her future self not to exist.
Einstein (a good friend of G\"odel) was himself seriously disturbed by
the discovery of CTCs~\cite{schilpp}.  Because the theory of P-CTCs
rely on post-selection, they provide self-consistent resolutions to
such paradoxes: anything that happens in a P-CTC can also happen in
conventional quantum mechanics with some probability.  Similarly, the
post-selected nature of P-CTCs allows the predictions and
retrodictions of the theory to be tested experimentally, even in the
absence of an actual general-relativistic closed timelike curve.

Time travel is a subject that has fascinated human beings for
thousands of years.  In the Hindu epic, the Mahabarata, for example,
King Revaita accepts an invitation to visit Brahma's palace.  Although
he stays for only a few days, when he returns to earth he finds that
many eons have passed.  The Japanese fisherman in the folk tale
Urashima Taro, having saved a sea turtle, is invited to the palace of
the sea-king; upon returning home discovers on the beach a crumbling
monument, centuries old, memorializing him.  The Gaelic hero Finn
McCool suffers a similar fate.  These stories also dwell on the
dangers of time travel.  Urashima Taro is given a magic box and told
not to open it.  Finn receives the gift of a magic horse and told not
to dismount.  When, inevitably, Taro opens the box, and Finn's toe
touches the ground, they instantaneously age and crumble into dust.

These tales involve time travel to the future.  Perhaps because
of the various paradoxes to which it gives rise, the concept
of travel to the past is a more recent invention.  Starting in
the late eighteenth century, a few narratives take a stab
at time travel to the past, the best known being Charles Dickens's
{\it A Christmas Carol,} and Mark Twain's {\it A Connecticut Yankee in
King Arthur's Court.}  The contemporary notion of time travel,
together with all its attendant paradoxes, did not come into
being until H.G. Wells' masterpiece, {\it The Time Machine},
which is also the first book to propose an actual device
that can be used to travel back and forth in time.

As frequently happens, scientific theories of time travel lagged
behind the fictional versions.  Although Einstein's theory of general
relativity implicitly allows travel to the past, it took several
decades before G\"odel proposed an explicit space-time geometry
containing closed timelike curves (CTCs).  The G\"odel universe
consists of a cloud of swirling dust, of sufficient gravitational
power to support closed timelike curves.  Later, it was realized that
closed timelike curves are a generic feature of highly curved,
rotating spacetimes: the Kerr solution for a rotating black hole
contains closed timelike curves within the black hole horizon; and
massive rapidly rotating cylinders typically are associated with
closed timelike curves \cite{Lanczos, STOCKUM, Bonnor}.  The topic of
closed timelike curves in general relativity continues to inspire
debate: Hawking's chronology protection postulate, for example,
suggests that the conditions needed to create closed timelike curves
cannot arise in any physically realizable spacetime~\cite{Hawking}.
For example, while Gott showed that cosmic string geometries can
contain closed timelike curves~\cite{Gott}, Deser {\it et al.}  showed
that physical cosmic strings cannot create CTCs from scratch
\cite{Deser, Carroll}.

At bottom, the behavior of matter is governed by the laws of quantum
mechanics.  Considerable effort has gone into constructing quantum
mechanical theories for closed timelike curves.  The initial efforts
to construct such theories involved path integral formulations of
quantum mechanics.  Hartle and Politzer pointed out that in the
presence of closed timelike curves, the ordinary correspondence
between the path-integral formulation of quantum mechanics and the
formulation in terms of unitary evolution of states in Hilbert space
breaks down~\cite{hartle, politzer}.  Morris {\it et al.}  explored
the quantum prescriptions needed to construct closed timelike curves
in the presence of wormholes, bits of spacetime geometry that, like
the handle of a coffee cup, `break off' from the main body of the
universe and rejoin it in the the past \cite{altri}.  Meanwhile,
Deutsch formulated a theory of closed timelike curves in the context
of Hilbert space, by postulating self-consistency conditions for the
states that enter and exit the closed timelike curve~\cite{deutsch}.

General relativistic closed timelike curves provide one
potential mechanism for time travel, but they need not provide
the only one.  Quantum mechanics supports a variety of counter-intuitive
phenomena which might allow time travel even in the absence
of a closed timelike curve in the geometry of spacetime.
One of the best-known versions of non-general relativistic
quantum versions of time travel comes from Wheeler, as
described by Feynman in his Nobel Prize lecture~\cite{FEYN}:

\par\noindent
`I received a telephone call one day at the graduate college at Princeton 
from Professor Wheeler, in which he said,

\par\noindent ``Feynman, I know why 
all electrons have the same charge and the same mass.''
 
\par\noindent ``Why?''

\par\noindent ``Because, they are all the same electron!''

And, then he explained on the telephone, 

\par\noindent ``Suppose that the world lines which we were 
ordinarily considering before in time and space - instead of only
going up in time were a tremendous knot, and then, when we cut through
the knot, by the plane corresponding to a fixed time, we would see
many, many world lines and that would represent many electrons, except
for one thing. If in one section this is an ordinary electron world
line, in the section in which it reversed itself and is coming back
from the future we have the wrong sign to the proper time - to the
proper four velocities - and that's equivalent to changing the sign of
the charge, and, therefore, that part of a path would act like a
positron.'' '

\bigskip\noindent As we will see, post-selected closed timelike
curves make up a precise physical theory which instantiates
Wheeler's whimsical idea.

The purpose of the current paper is to provide a unifying description
of closed timelike curves in quantum mechanics.  We start from the
prescription that time travel effectively represents a communication
channel from the future to the past.  Quantum time travel, then,
should be described by a quantum communication channel to the past.  A
well-known quantum communication channel is given by quantum
teleportation, in which shared entanglement combined with quantum
measurement and classical communication allows quantum states to be
transported between sender and receiver.  We show that if quantum
teleportation is combined with post-selection, then the result is a
quantum channel to the past.  The entanglement occurs between the
forward- and backward- going parts of the curve, and post-selection
replaces the quantum measurement and obviates the need for classical
communication, allowing time travel to take place.  The resulting
theory allows a description both of the quantum mechanics of general
relativistic closed timelike curves, and of Wheeler-like quantum time
travel in ordinary spacetime.

As described in previous work~\cite{loops}, the notion that
entanglement and projection can give rise to closed timelike curves to
has arisen independently in a variety of contexts.  This combination
lies at the heart of the Horowitz-Maldacena model for information
escape from black holes~\cite{HM, Yurtsever, Gottesman, Lloyd1}, and
Gottesmann and Preskill note in passing that this mechanism might be
used for time travel~\cite{Gottesman}.  Pegg explored the use of a
related mechanism for `probabilistic time machines'~\cite{pegg}.
Bennett and Schumacher have explored similar notions in unpublished
work~\cite{benschu}.  Ralph suggests using teleportation for time
traveling, although in a different setting, namely, displacing the
entangled resource in time~\cite{ralph3}.  Svetlichny describes
experimental techniques for investigating quantum travel based on
entanglement and projection~\cite{SVE}. Chiribella {\em eet al.}
cosider this mechanism while analyzing extensions to the quantum
computational model~\cite{mauro1}. Brukner {\em et al.}  have analyzed
probabilistic teleportation (where only the cases in which the Bell
measurement yields the desired result are retained) as a computational
resource in~\cite{zeil}.

The outline of the paper follows. In Sec.~\ref{s:pctcs} we describe
P-CTCs and Deutsch's mechanism in detail, emphasizing the differences
between the two approaches. Then, in Sec.~\ref{s:pathint} we relate
P-CTCs to the path-integral formulation of quantum mechanics.  This
formulation is particularly suited for the description of quantum
field theory in curved spacetime~\cite{birrel}, and has been used
before to provide quantum descriptions of closed timelike
curves~\cite{novikov, friedman, friedman1, hartle, politzer, boulware,
  politzer2, diaz}.  Our proposal is consistent with these
path-integral approaches. In particular, the path-integral description
of fermions using Grassmann fields given by Politzer~\cite{politzer}
yields a dynamical description which coincides with ours for systems
of quantum bits. Other descriptions, such as Hartle's~\cite{hartle},
are more difficult to compare as they do not provide an explicit
prescription to calculate the details of the dynamics of the
interaction with systems inside closed timelike curves. In any case,
their general framework is consistent with our derivations. By
contrast, Deutsch's CTCs are not compatible with the Politzer
path-integral approach, and are analyzed by him on a different
footing~\cite{politzer}. Indeed, suppose that the path integral is
performed over classical paths which agree both at the entrance to--
and at the exit from-- the CTC, so that $x$-in, $p$-in are the same as
$x$-out, $p$-out. Similarly, in the Grassmann case, suppose that
spin-up along the $z$ axis at the entrance emerges as spin-up along
the $z$ axis at the exit. Then, the quantum version of the CTC must
exhibit the same perfect correlation between input and output. But, as
the grandfather paradox experiment~\cite{loops} shows, Deutsch's CTCs
need not exhibit such correlations: spin-up in is mapped to spin-down
out (although the overall quantum state remains the same). By
contrast, P-CTCs exhibit perfect correlation between in- and out-
versions of all variables. Note that a quantum-field theoretical
justification of Deutsch's solution is proposed in~\cite{ralph1,
  ralph2} and is based on introducing additional Hilbert subspaces for
particles and fields along the geodesic: observables at different
points along the geodesic commute because they act on different
Hilbert spaces.

The path-integral formulation also shows that using P-CTCs it is
impossible to assign a well defined state to the system in the CTC.
This is a natural requirement (or, at least, a desirable property),
given the cyclicity of time there. In contrast, Deutsch's consistency
condition~\eqref{conscon} is explicitly built to provide a
prescription for a definite quantum state $\rho_{CTC}$ of the system
in the CTC.

In Sec.~\ref{s:general} we go beyond the path-integral formulation and
provide the dynamical evolution formulas in the context of generic
quantum mechanics (the Hilbert-space formulation). Namely, we treat
the CTC as a generic quantum transformation, where the transformed
system emerges at a previous time ``after'' eventually interacting
with some chronology-respecting systems. In this framework we obtain
the explicit prescription of how to calculate the nonlinear evolution
of the state of the system in the chronology-respecting part of the
spacetime. This nonlinearity is exactly of the form that previous
investigations (e.g.~Hartle's~\cite{hartle}) have predicted.

In Sec.~\ref{s:generalrel} we consider time travel situations that are
independent from general-relativistic CTCs. We then conclude in
Sec.~\ref{s:comp} with considerations on the computational power of
the different models of CTCs.

\section{P-CTCs and Deutsch's CTCs}\labell{s:pctcs}

Any quantum theory of gravity will have to propose a prescription to
deal with the unavoidable~\cite{hartle} nonlinearities that plague
CTCs. This requires some sort of modification of the dynamical
equations of motions of quantum mechanics that are always linear.
Deutsch in his seminal paper~\cite{deutsch} proposed one such
prescription, based on a self-consistency condition referred to the
state of the systems inside the CTC.  Deutsch's theory has recently
been critiqued by several authors as exhibiting self-contradictory
features~\cite{bennett, ralph, ralph1, ralph2}.  By contrast, although
any quantum theory of time travel quantum mechanics is likely to yield
strange and counter-intuitive results, P-CTCs appear to be less
pathological~\cite{loops}.  They are based on a different
self-consistent condition that states that self-contradictory events
do not happen (Novikov principle~\cite{novikov}). Pegg points out that
this can arise because of destructive interference of
self-contradictory histories~\cite{pegg}. Here we further compare
Deutsch's and post-selected closed timelike curves, and give an
in-depth analysis of the latter, showing how they can be naturally
obtained in the path-integral formulation of quantum theory and
deriving the equations of motions that describe the interactions with
CTCs.  As noted, in addition to general-relativistic CTCs, our
proposed theory can also be seen as a theoretical elaboration of
Wheeler's assertion to Feynman that `an electron is a positron moving
backward in time'~\cite{FEYN}.  In particular, any quantum theory
which allows the nonlinear process of postselection supports time
travel even in the absence of general-relativistic closed timelike
curves.

The mechanism of P-CTCs~\cite{loops} can be summarized by saying that
they behave exactly as if the initial state of the system in the P-CTC
were in a maximal entangled state (entangled with an external
purification space) and the final state were post-selected to be in
the same entangled state. When the probability amplitude for the
transition between these two states is null, we postulate that the
related event does not happen (so that the Novikov
principle~\cite{novikov} is enforced). By contrast, Deutsch's CTCs are
based on imposing the consistency condition
\begin{eqnarray}
\rho_{CTC}=\mbox{Tr}_A[U(\rho_{CTC}\otimes\rho_A)U^\dag],
\labell{conscon}\;
\end{eqnarray}
where $\rho_{CTC}$ is the state of the system inside the closed
timelike curve, $\rho_A$ is the state of the system outside (i.e.~of
the chronology-respecting part of spacetime), $U$ is the unitary
transformation that is responsible for eventual interactions among the
two systems, and where the trace is performed over the
chronology-respecting system. The existence of a state $\rho$ that
satisfies~\eqref{conscon} is ensured by the fact that any
completely-positive map of the form ${\cal
  L}[\rho]=\mbox{Tr}_A[U(\rho\otimes\rho_A)U^\dag]$ always has at
least one fixed point $\rho$ (or, equivalently, one eigenvector $\rho$
with eigenvalue one). If more than one state $\rho_{CTC}$ satisfies
the consistency condition~\eqref{conscon}, Deutsch separately
postulates a ``maximum entropy rule'', requesting that the maximum
entropy one must be chosen.  Note that Deutsch's formulation assumes that
the state exiting the CTC in the past is completely uncorrelated
with the chronology-preserving variables at that time:
the time-traveler's `memories' of events in the future are
no longer valid.

\begin{figure}[h!]
\begin{center}
\epsfxsize=.55\hsize\epsffile{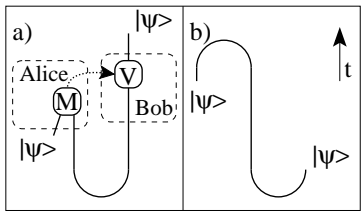} 
\end{center}
\caption{Description of closed timelike curves through teleportation.
  a) Conventional teleportation: Alice and Bob start from a maximally
  entangled state shared among them represented by ``$\bigcup$''.
  Alice performs a Bell measurement M on her half of the shared state
  and on the unknown state $|\psi\rangle$ she wants to transmit.  This
  measurement tells her which entangled state the two systems are in.
  She then communicates (dotted line) the measurement result to Bob
  who performs a unitary V on his half of the entangled state,
  obtaining the initial unknown state $|\psi\rangle$. b) Post-selected
  teleportation: the system in state $|\psi\rangle$ and half of the
  Bell state ``$\bigcup$'' are projected onto the same Bell state
  ``$\bigcap$''.  This means that the other half of the Bell state is
  projected into the initial state of the system $|\psi\rangle$ even
  {\em before} this state is available.
  \label{f:teleport}}
\end{figure}

The primary conceptual difference between Deutsch's CTCs and P-CTCs
lies in the self-consistency condition imposed.  Consider a
measurement that can be made either on the state of the system as it
enters the CTC, or on the state as it emerges from the CTC.  Deutsch
demands that these two measurements yield the same statistics for the
CTC state alone: that is, the density matrix of the system as it
enters the CTC is the same as the density matrix of the system as it
exits the CTC.  By contrast, we demand that these two measurements
yield the same statistics for the CTC state {\it together with its
  correlations with any chronology preserving variables}.  It is this
demand that closed timelike curves respect both statistics for the
time-traveling state together with its correlations with other
variables that distinguishes P-CTCs from Deutsch's CTCs.  The fact
that P-CTCs respect correlations effectively enforces the Novikov
principle~\cite{novikov}, and, as will be seen below, makes P-CTCs
consistent with path-integral approaches to CTCs.

The connection between P-CTCs and teleportation~\cite{teleportation}
is illustrated (see Fig.~\ref{f:teleport}) with the following simple
example that employs qubits (extensions to higher dimensional systems
are straightforward).  Suppose that the initial Bell state is
$|\Psi^{(-)}\rangle\propto |01\rangle-|10\rangle$ (but any maximally
entangled Bell state will equivalently work), and suppose that the
initial state of the system entering the CTC is $|\psi\rangle$.  Then
the joint state of the three systems (system~1 entering the CTC,
system~2 emerging from the CTC, and system~3, its purification) is
given by $|\psi\rangle_1|\Psi^{(-)}\rangle_{23}$.  These three systems
are denoted by the three vertical lines of Fig.~\ref{f:teleport}b. It
is immediate to see that this state can be also written as
  \begin{eqnarray}
&&    (-|\Psi^{(-)}\rangle_{13}|\psi\rangle_2-
|\Psi^{(+)}\rangle_{13}\sigma_z|\psi\rangle_2+\nonumber\\&&
|\Phi^{(-)}\rangle_{13}\sigma_x|\psi\rangle_2+
i|\Phi^{(+)}\rangle_{13}\sigma_y|\psi\rangle_2)/2
\labell{telep}\;,
\end{eqnarray}
where $|\Psi^{(\pm)}\rangle\propto|01\rangle\pm|10\rangle$ and
$|\Phi^{(\pm)}\rangle\propto|00\rangle\pm|11\rangle$ are the four
states in a Bell basis for qubit systems and $\sigma_\alpha$s are the
three Pauli matrices. Eq.~\eqref{telep} is equivalent to Eq.~(5) of
Ref.~\cite{teleportation}, where the extension to higher dimensional
systems is presented (the extension to infinite dimensional systems is
presented in~\cite{braunstein}). It is immediate to see that, if the
system~1 entering the CTC together with the purification system~3 are
post-selected to be in the same Bell state $|\Psi^{(-)}\rangle_{13}$
as the initial one, then only the first term of Eq.~\eqref{telep}
survives. Apart from an inconsequential minus sign, this implies that
the system~2 emerging from the CTC is in the state $|\psi\rangle_2$,
which is exactly the same state of the system that has entered
(rather, will enter) the CTC.

It seems that, based on what is currently known on these two
approaches, we cannot conclusively choose P-CTCs over Deutsch's, or
{\it vice versa}.  Both arise from reasonable physical assumptions and
both are consistent with different approaches to reconciling quantum
mechanics with closed timelike curves in general relativity.  A final
decision on which of the two is ``actually the case'' may have to be
postponed to when a full quantum theory of gravity is derived (which
would allow to calculate from first principles what happens in a CTC)
or when a CTC is discovered that can be tested experimentally.
However, because of the huge recent interest on CTCs in physics and in
computer science (e.g.  see~\cite{bacon, bennett, francesi, aar1,
  aar2, Brun1, ralph}), it is important to point out that there are
reasonable alternatives to the leading theory in the field.  We also
point out that our post-selection based description of CTCs seems to
be less pathological than Deutsch's: for example P-CTCs have less
computational power and do not require to separately postulate a
``maximum entropy rule''~\cite{loops}.  Therefore, they are in some
sense preferable, at least from an Occam's razor perspective.
Independent of such questions of aesthetic preference, as we will now
show, P-CTCs are consistent with previous path integral formulations
of closed timelike curves, whereas Deutsch's CTCs are not.



\section{P-CTCs and path integrals}\labell{s:pathint}

Path integrals~\cite{feynm1,feynm2} allow one to calculate the
transition amplitude for going from an initial state $|I\rangle$ to a
final state $|F\rangle$ as an integral over paths of the action, i.e.
\begin{widetext}\begin{eqnarray}
\langle F|\exp(-\tfrac i\hbar H\tau)|I\rangle=
\int_{-\infty}^{+\infty} dx\: dy\; I(x)\;
F^*(y)\int_x^y{\cal D}x(t)\exp[\tfrac i\hbar S],\mbox{ where }
S=\int_0^{\tau} dtL(x,\dot x)
\labell{feyn}\;,
\end{eqnarray}
and where $L={T}-V$ is the Lagrangian and $S$ is the action, $I(x)$
and $F(x)$ are the position representations of $|I\rangle$ and
$|F\rangle$ respectively (i.e.~$|I\rangle=\int dx\:I(x)|x\rangle$),
and the paths in the integration over paths all start in $x$ and end
in $y$. Of course in this form it is suited only to describing the
dynamics of a particle in space (or a collection of particles). It
will be extended to other systems in the next section.

In order to add a CTC, we first divide the spacetime into two parts,
\begin{eqnarray}
  \langle F|_C\langle F'|\exp(-\tfrac i\hbar H\tau)|I\rangle|I'\rangle_C=
  \int_{-\infty}^{+\infty} dx\: dx'dy\:dy'\; I(x)\;I'(x')\;
  F^*(y)\;
  F'^*(y')  \int_{x,x'}^{y,y'}{\cal D}x(t)\exp[\tfrac i\hbar S]
\labell{feyn2}\;,
\end{eqnarray}
The ``conventional'' strategy to deal with CTCs using path integrals
is to send the system $C$ to a prior time unchanged (i.e. with the
same values of $x,\dot x$), while the other system (the
chronology-respecting one) evolves normally. This is enforced by
imposing periodic boundary conditions on the CTC boundaries. Namely,
the probability amplitude for the chronology-respecting system is
\begin{eqnarray}
  \langle F|\exp(-\tfrac i\hbar H\tau)|I\rangle=
  \int_{-\infty}^{+\infty} dx\: dx'dy\:dy'\; I(x)\;
  F^*(y)\;\delta(x'-y') \int_{x,x'}^{y,y'}{\cal D}x(t)\exp[\tfrac i\hbar S]
\labell{feyn3}\;,
\end{eqnarray}
where the $\delta$-function ensures that the initial and final
boundary conditions in the CTC system are the same. Note that we have
removed $I'(x')$ and $F'(y')$, but we are coherently adding all
possible initial and final conditions (through the $x'$ and $y'$
integrals). This implies that it is not possible to assign a definite
state to the system inside a CTC: all possible states of the system
(except possibly forbidden ones) are compatible with such boundary
conditions. Note also that the boundary conditions of
Eq.~\eqref{feyn3} have previously appeared in the literature
(e.g.~see~\cite{politzer2} and, in the classical context, in the
seminal paper~\cite{thorne}).

\vskip1\baselineskip To show that Eq.~\eqref{feyn3} is
the same formula that one obtains using post-selected teleportation,
we have to calculate $\langle F|_C\langle\Psi|\exp(-\tfrac i\hbar
H\tau)\otimes\openone|I\rangle|\Psi\rangle_C$, where
$|\Psi\rangle\propto\int dx|xx\rangle$ is a maximally entangled EPR state~\cite{epr}
and where the Hamiltonian acts only on the system and on the first of
the two Hilbert spaces of $|\Psi\rangle$. Use Eq.~\eqref{feyn2} for
the system and for the first Hilbert space of $|\Psi\rangle$ to obtain
\begin{eqnarray}
&&  \langle F|_C\langle\Psi|\exp(-\tfrac i\hbar
  H\tau)\otimes\openone|I\rangle|\Psi\rangle_C=
  \int_{-\infty}^{+\infty} dx\: dx'dy\:dy'dz\:dz'\; I(x)\;F^*(y)\;
  \delta(x'-z)\;
  \delta(y'-z')  \langle z|\openone|z'\rangle \nonumber\\&&\times
  \int_{x,x'}^{y,y'}{\cal D}x(t)\exp[\tfrac i\hbar S]
= \int^{\infty}_{-\infty} dx dx^{\prime} dy dy^{\prime} I (x)
F^{\ast}
(y)\;
\delta(x'-y') \int^{y, y^{\prime}}_{x, x^{\prime}} {\cal D} x(t) \exp \left[
  \frac{i}{\hbar} S \right] 
\labell{tp}\;,
\end{eqnarray}\end{widetext}
where we have used the position representation $|\Psi\rangle=\int
dy\:dz\:\delta(y-z)|y\rangle|z\rangle$.  Eq.~\eqref{tp} is clearly
equal to Eq.~\eqref{feyn3} since $\langle z|z'\rangle=\delta(z-z')$.
Note that this result is independent of the particular form of the EPR
state $|\Psi\rangle$ as long as it is maximally entangled in position
(and hence in momentum).

All the above discussion holds for initial and final pure states.
However, the extension to mixed states in the path-integral
formulation is straightforward: one only needs to employ appropriate
purification spaces~\cite{yutaka,mauro}. The formulas then
reduce to the previous ones.\togli{, namely~Eqs.~\eqref{feyn3} and
  \eqref{feyn4}.  }

Here we briefly comment on the two-state vector formalism of quantum
mechanics~\cite{weak1,weak2}. It is based on post-selection of the
final state and on renormalizing the resulting transition amplitudes:
it is a time-symmetrical formulation of quantum mechanics in which not
only the initial state, but also the final state is specified. As
such, it shares many properties with our post-selection based
treatment of CTCs. In particular, in both theories it is impossible to
assign a definite quantum state at each time: in the two-state
formalism the unitary evolution forward in time from the initial state
might give a different mid-time state with respect to the unitary
evolution backward in time from the final state. Analogously, in a
P-CTC, it is impossible to assign a definite state to the CTC system
at any time, given the cyclicity of time there. This is evident, for
example, from Eq.~\eqref{feyn3}: in the CTC system no state is
assigned, only periodic boundary conditions.  Another aspect that the
two-state formalism and P-CTCs share is the nonlinear renormalization
of the states and probabilities. In both cases this arises because of
the post-selection. In addition to the two-state formalism, our
approach can also be related to weak values~\cite{weak, weak1}, since
we might be performing measurements between when the system emerges
from the CTC and when it re-enters it.  Considerations analogous to
the ones presented above apply. It would be a mistake, however, to
think that the theory of post-selected closed timelike curves in some
sense requires or even singles out the weak value theory.  Although
the two are compatible with each other, the theory of P-CTCs is
essentially a `free-standing' theory that does not give preference to
one interpretation of quantum mechanics over another.
\section{General systems}\labell{s:general}
The formula~\eqref{feyn3} was derived in the path-integral formulation
of quantum mechanics, but it can be easily extended to generic quantum
evolution.

We start by recalling the usual Kraus decomposition of a generic
quantum evolution (that can describe the evolution of both isolated
and open systems). It is given by
\begin{eqnarray}
&&{\cal L}[\rho]=\mbox{Tr}_E[U(\rho\otimes|e\rangle\langle e|)U^\dag]=
\sum_i\langle i|U|e\rangle\:\rho\:\langle
e|U^\dag|i\rangle\nonumber\\&&
=\sum_iB_i\rho B_i^\dag
\labell{cp}\;,
\end{eqnarray}
where $|e\rangle$ is the initial state of the environment (or,
equivalently, of a putative abstract purification space), $U$ is the
unitary operator governing the interaction between system initially in
the state $\rho$ and environment, and $B_i\equiv\langle i|U|e\rangle$
is the Kraus operator. In contrast, the nonlinear evolution of our
post-selected teleportation scheme is given by
\begin{eqnarray}
&&{\cal N}[\rho]=\mbox{Tr}_{EE'}
\Big[(U\otimes\openone_{E'})
\left(\rho\otimes|\Psi\rangle\langle\Psi|\right)
\left(U^\dag\otimes\openone_{E'}\right)\nonumber\\&&\times
\left(\openone\otimes|\Psi\rangle\langle\Psi|\right)\Big]=
\sum_{l,j}\langle l|U|l\rangle\:\rho\:\langle j|U^\dag|j\rangle=
C\rho C^\dag
\labell{nl},
\end{eqnarray}
where $C\equiv$Tr$_{CTC}[U]$ and
$|\Psi\rangle\propto\sum_i|i\rangle_E|i\rangle_{E'}$ (or any other
maximally entangled state, which would give the same result).
Obviously, the evolution in \eqref{nl} is nonlinear (because of the
post-selection), so one has to renormalize the final state: ${\cal
  N}[\rho]\to {\cal N}[\rho]/$Tr${\cal N}[\rho]$. In other words,
according to our approach, a chronology-respecting system in a state
$\rho$ that interacts with a CTC using a unitary $U$ will undergo the
transformation
\begin{eqnarray}
{\cal
  N}[\rho]=\frac{C\;\rho\;C^\dag}
{\mbox{Tr}[C\;\rho\;C^\dag]}
\labell{evol}\;,
\end{eqnarray}
where we suppose that the evolution does not happen if
$C\equiv$Tr$_{CTC}[U]=0$. The comparison with~\eqref{cp} is
instructive: there the non-unitarity comes from the inaccessibility of
the environment. Analogously, in~\eqref{evol} the non-unitarity comes
from the fact that, after the CTC is closed, for the
chronology-respecting system it will be forever inaccessible. The
nonlinearity of~\eqref{evol} is more difficult to interpret, but is
connected with the periodic boundary conditions in the CTC.  Note that
this general evolution equation~\eqref{evol} is consistent with
previous derivations based on path integrals. For example, it is
equivalent to Eq.~(4.6) of Ref.~\cite{hartle} by Hartle. However, in
contrast to here, the actual form of the evolution operators $C$ is
not provided there. As a further example, consider Ref.~\cite{politzer},
where Politzer derives a path integral approach of CTCs for qubits,
using Grassmann fields. His Eq.~(5) is compatible with Eq.~\eqref{nl}.
He also derives a nonunitary evolution that is consistent with
Eq.~\eqref{evol} in the case in which the initial state is pure.  In
particular, this implies that, also in the general qudit case, our
post-selected teleportation approach gives the same result one would
obtain from a specific path-integral formulation. In addition, it has
been pointed out many times before (e.g.~see~\cite{friedman, cassidy})
that when quantum fields inside a CTC interact with external fields,
linearity and unitarity is lost. It is also worth to notice that there
have been various proposals to restore unitarity by modifying the
structure of quantum mechanics itself or by postulating an
inaccessible purification space that is added to uphold
unitarity~\cite{anderson, wells}.

The evolution~\eqref{evol} coming from our approach is to be compared
with Deutsch's evolution,
\begin{eqnarray}
&&{\cal D}[\rho]=\mbox{Tr}_{CTC}[U(\rho_{CTC}\otimes\rho)U^\dag],\mbox{
  where }\nonumber\\&&\rho_{CTC}=\mbox{Tr}_A[U(\rho_{CTC}\otimes\rho)U^\dag]
\labell{consconds}\;
\end{eqnarray}
satisfies the consistency condition. \togli{As we have shown in the
  previous section, also this evolution can be derived in a path
  integral formulation, by assigning different boundary conditions.}
The direct comparison of Eqs.~\eqref{evol} and~\eqref{consconds}
highlights the differences in the general prescription for the
dynamics of CTCs of these two approaches.

Even though the results presented in this section are directly
applicable only to general finite-dimensional systems, the extension
to systems living in infinite-dimensional separable Hilbert spaces
seems conceptually straightforward, although mathematically involved.

In his path-integral formulation of CTCs, Hartle notes that CTCs
might necessitate abandoning not only unitarity and linearity,
but even the familiar Hilbert space formulation of quantum
mechanics \cite{hartle}.  Indeed, the fact that the state of a system
at a given time can be written as the tensor product states
of subsystems relies crucially on the fact that operators
corresponding to spacelike separated regions of spacetime
commute with each other.  When CTCs are introduced, the notion
of `spacelike' separation becomes muddied.  The formulation
of closed timelike curves in terms of P-CTCs shows, however,
that the Hilbert space structure of quantum mechanics can
be retained. 

\section{Time travel in the absence of general-relativistic
  CTCs}\labell{s:generalrel}

Although the theory of P-CTCs was developed to address the question of
quantum mechanics in general-relativistic closed timelike curves, it
also allows us to address the possibility of time travel in other
contexts.  Essentially, any quantum theory that allows the nonlinear
process of projection onto some particular state, such as the
entangled states of P-CTCs, allows time travel even when no spacetime
closed timelike curve exists.  Indeed, the mechanism for such time
travel closely follows Wheeler's famous telephone call above.
Non-general-relativistic P-CTCs can be implemented by the creation of
and projection onto entangled particle-antiparticle pairs.  Such a
mechanism is just what is used in our experimental tests of P-CTCs
~\cite{loops}: although projection is a non-linear process that cannot
be implemented deterministically in ordinary quantum mechanics, it can
easily be implemented in a probabilistic fashion.  Consequently, the
effect of P-CTCs can be tested simply by performing quantum
teleportation experiments, and by post-selecting only the results that
correspond to the desired entangled-state output.

If it turns out that the linearity of quantum mechanics is only
approximate, and that projection onto particular states does in fact
occur -- for example, at the singularities of black holes~\cite{HM,
  Yurtsever, Gottesman, Lloyd1} -- then it might be possible to
implement time travel even in the absence of a general-relativistic
closed timelike curve.  The formalism of P-CTCs shows that such
quantum time travel can be thought of as a kind of quantum tunneling
backwards in time, which can take place even in the absence of a
classical path from future to past.

\section{Computational power of CTCs}\labell{s:comp}

It has been long known that nonlinear quantum mechanics potentially
allows the rapid solution of hard problems such as NP-complete
problems~\cite{abrams}.  The nonlinearities in the quantum mechanics
of closed timelike curves is no exception~\cite{aar1, aar2, Brun1}.
Aaronson and Watrous have shown quantum computation with Deutsch's
closed timelike curves allows the solution of any problem in PSPACE,
the set of problems that can be solved using polynomial space
resources~\cite{aar1}.  Similarly, Aaronson has shown that quantum
computation combined with post-selection allows the solution of any
problem in the computational class PP, probabilistic polynomial
time(where a probabilistic polynomial Turing machine accepts with
probability $\tfrac 12$ if and only if the answer is ``yes.'').
Quantum computation with post-selection explicitly allows P-CTCs, and
P-CTCs in turn allow the performance of any desired post-selected
quantum computation.  Accordingly, quantum computation with P-CTCs can
solve any problem in PP, including NP-complete problems.  Since the
class PP is thought to be strictly contained in PSPACE, quantum
computation with P-CTCs is apparently strictly less powerful than
quantum computation with Deutsch's CTCs.

In the case of quantum computating with Deutschian CTCs, Bennett {\it
  et al.}~\cite{bennett} have questioned whether the notion of
programming a quantum computer even makes sense.  Ref.~\cite{bennett}
notes that in Deutsch's closed timelike curves, the nonlinearity
introduces ambiguities in the definition of state preparation: as is
well-known in nonlinear quantum theories, the result of sending {\it
  either} the state $|\psi\rangle$ through a closed-timelike curve
{\it or} the state $|\phi\rangle$ is no longer equivalent to sending
the mixed state $(1/2)( |\psi\rangle\langle\psi| + |\phi\rangle\langle
\phi|)$ through the curve.  The problem with computation arises
because, as is clear from our grandfather-paradox
circuit~\cite{loops}, Deutsch's closed timelike curves typically break
the correlation between chronology preserving variables and the
components of a mixed state that enters the curve: the component that
enters the CTC as $|0\rangle$ can exit the curve as $|1\rangle$, even
if the overall mixed state exiting the curve is the same as the one
that enters.  Consequently, Bennett {\it et al.} argue, the programmer
who is using a Deutschian closed timelike curve as part of her quantum
computer typically finds the output of the curve is completely
decorrelated from the problem she would like to solve: the curve emits
random states.

\begin{figure}
\begin{center}
\epsfxsize=.4\hsize\epsffile{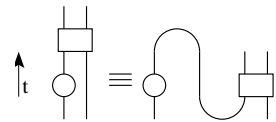} 
\end{center}
\caption{
   Closed timelike loops can
 collapse the time-depth of any circuit to one,
 allowing to compute any problem not merely efficiently, but
instantaneously. 
 \label{f:comput}}
\end{figure}

In contrast, because P-CTCs are formulated explicitly to retain
correlations with chronology preserving curves, quantum computation
using P-CTCs do not suffer from state-preparation ambiguity.  That is
not so say that P-CTCs are computationally innocuous: their nonlinear
nature typically renormalizes the probability of states in an input
superposition, yielding to strange and counter-intuitive effects.  For
example, any CTC can be used to compress any computation to depth one,
as shown in Fig.~\ref{f:comput}. Indeed, it is exactly the ability of
nonlinear quantum mechanics to renormalize probabilities from their
conventional values that gives rise to the amplification of small
components of quantum superpositions that allows the solution of hard
problems.  Not least of the counter-intuitive effects of P-CTCs is
that they could still solve hard computational problems with ease!
The `excessive' computational power of P-CTCs is effectively an
argument for why the types of nonlinearities that give rise to P-CTCs,
if they exist, should only be found under highly exceptional
circumstances such as general-relativistic closed timelike curves or
black-hole singularities.

\section{Conclusions}\labell{s:concl}

This paper reviewed quantum mechanical theories for time travel,
focusing on the theory of P-CTCs~\cite{loops}.  Our purpose in
presenting this work is to make precise the similarities and
differences between varying quantum theories of time travel.  We
summarize our findings here.

We have extensively argued that P-CTCs are physically inequivalent to
Deutsch's CTCs. In Sec.~\ref{s:pathint} we showed that P-CTCs are
compatible with the path-integral formulation of quantum mechanics.
This formulation is at the basis of most of the previous analysis of
quantum descriptions of closed time-like curves, since it is
particularly suited to calculations of quantum mechanics in curved
space time.  P-CTCs are reminiscent of, and consistent with, the
two-state-vector and weak-value formulation of quantum mechanics.  It
is important to note, however, that P-CTCs do not in any sense require
such a formulation. Then, in Sec.~\ref{s:general} we extended our
analysis to general systems where the path-integral formulation may
not always be possible and derived a simple prescription for the
calculation of the CTC dynamics, namely Eq.~\eqref{evol}. In this way
we have performed a complete characterization of P-CTC in the most
commonly employed frameworks for quantum mechanics, with the exception
of algebraic methods (e.g.~see~\cite{yurtsever}).

In Sec.~\ref{s:generalrel} we have argued that, as Wheeler's picture
of positrons as electrons moving backwards in time suggests, P-CTCs
might also allow time travel in spacetimes without
general-relativistic closed timelike curves.  If nature somehow
provides the nonlinear dynamics afforded by final-state projection,
then it is possible for particles (and, in principle, people) to
tunnel from the future to the past.

Finally, in Sec.~\ref{s:comp} we have seen that P-CTCs are
computationally very powerful, though less powerful than the
Aaronson-Watrous theory of Deutsch's CTCs.

Our hope in elaborating the theory of P-CTCs is that this theory may
prove useful in formulating a quantum theory of gravity, by providing
new insight on one of the most perplexing consequences of general
relativity, i.e., the possibility of time-travel.


\end{document}